\documentclass[twocolumn,aps,superscriptaddress,pre]{revtex4}
\usepackage{amsmath,amsfonts,amssymb,graphicx}
\usepackage{psfrag}
\usepackage{wrapfig}
\usepackage{subfigure}
\usepackage{makeidx}
\usepackage{bm}
\usepackage{epsf}
\usepackage{multirow}
\usepackage[colorlinks,linkcolor=green,urlcolor=blue,citecolor=blue]{hyperref}
\DeclareMathOperator{\sech}{sech}

\begin{document}
\title{A direct derivation of the dark soliton excitation energy}

\author{Li-Chen Zhao}
\affiliation{School of Physics, Northwest University, Xi'an 710127, China}
\affiliation{Shaanxi Key Laboratory for Theoretical Physics Frontiers, Xi'an 710127, China}

\author{Yan-Hong Qin}
\affiliation{School of Physics, Northwest University, Xi'an 710127, China}
\affiliation{Shaanxi Key Laboratory for Theoretical Physics Frontiers, Xi'an 710127, China}

\author{Wenlong Wang}
\email{wenlongcmp@scu.edu.cn}
\affiliation{College of Physics, Sichuan University, Chengdu 610065, China}

\author{Zhan-Ying Yang}
\affiliation{School of Physics, Northwest University, Xi'an 710127, China}
\affiliation{Shaanxi Key Laboratory for Theoretical Physics Frontiers, Xi'an 710127, China}

 %%%%%%%%%%%%%%%%%%%%%%%%%%%%%%%%%%%%%%%%%%%%%%%%%﹛
\date{\today}
\begin{abstract}
Dark solitons are common topological excitations in a wide array of nonlinear waves. The dark soliton excitation energy, crucial for exploring dark soliton dynamics, is necessarily calculated in a renormalized form due to its existence on a finite background. Despite its tremendous importance and success, the renormalized energy form was firstly only suggested with no detailed derivation, and was then ``derived'' in the grand canonical ensemble. In this work, we revisit this fundamental problem and provide an alternative and intuitive derivation of the energy form from the fundamental field energy by utilizing a limiting procedure that conserves number of particles. Our derivation yields the same result, putting therefore the dark soliton energy form on a solid basis.
\end{abstract}
\maketitle

\section{Introduction}
Dark solitons are fundamental topological excitations of numerous nonlinear waves \cite{darks}, and have received much attention in the past several decades. A dark soliton is a spatially localized density ``dip" on top of a finite background, accompanied with a phase step through the dip \cite{Kivshar0}; the phase step is $\pi$ if the dark soliton is stationary.
Dark solitons exist in numerous physical systems, such as liquids \cite{liquids}, thin magnetic films \cite{films}, optical media \cite{opt1,opt2,opt3}, and Bose-Einstein condensates \cite{darkex1,darkex2} among others. One theme of research in this broad field is to study the effective nonlinear dynamics of the dark solitons, highlighting the particle aspect of these solitary waves. In order to discuss the effective mass and related kinetic dynamics of dark solitons \cite{Kivshar1,Kivshar2,Busch,Lewenstein,Konotop,Brazhnyi,Pelinovsky,PNAS,Hurst,Serkin}, it is essential to characterize the excitation energy of a dark soliton. Importantly, this energy also serves as a starting point for investigating dynamics of dark soliton filaments and surfaces in higher dimensions \cite{Wang,Cisneros}.

It is not straightforward to extract the dark soliton excitation energy as one needs to properly subtract the contribution of the finite background. This is in contrast to a bright soliton, where there is a density ``hump'' on a zero background. The renormalized dark soliton energy was firstly suggested in 1994 \cite{Kivshar1}, and is proven to be very effective, yielding excellent agreement between theory and experiment \cite{Weller,darkex,Frantzeskakis}. The dark soliton energy was also ``derived'' from the ``grand canonical energy" or free energy \cite{Busch,Konotop}. This formalism works indirectly with a constant chemical potential rather than a constant number of particles. However, these pioneering arguments and calculations are not entirely rigorous or clear.

The main purpose of this work is to provide an alternative and importantly intuitive derivation of the dark soliton excitation energy with details. We focus on a finite domain and calculate the excitation energy using the fundamental field energy definition by keeping the number of particles conserved. By taking the infinite limit of the domain, our result remarkably recovers to the well-known expression of the dark soliton excitation energy. Our derivation is therefore conceptually clear and natural compared with the free energy calculation, since the Hamiltonian of interest (see the next section) does not actually admit particle number variation. Our derivation, due to the finite domain and the limiting procedure, is technically more complicated but nevertheless straightforward and readily tractable.

This paper is organized as follows. In Sec.~\ref{mm}, we introduce
the model and the method. Next, we present our analytical results and discussions in Sec.~\ref{results}.
Finally, our conclusions and open
problems for future consideration are given in Sec.~\ref{conclusion}.

\section{Model and method}
\label{mm}
We now focus on a specific system of Bose-Einstein condensate for ease of discussion, although our following results are also valid for many other systems described by the same equation, particularly for optics. Here, we work in the quasi-1D regime, as we are interested in dark soliton states \cite{darkex1,darkex2}. In the framework of the mean-field theory, the dynamics of a repulsive cigar-shaped condensate at sufficiently low temperatures can be described by the following dimensionless Gross-Pitaevskii (GP) equation:
\begin{eqnarray}
i \psi_t&=& -\frac{1}{2} \psi_{xx}+ |\psi|^2 \psi,
\end{eqnarray}
where $\psi(x,t)$ is the macroscopic wavefunction. It is well-known that the system conserves among others the field energy $H = \int_{-\infty}^{\infty} [ \psi^{*}(-\frac{1}{2}\partial_{x}^2) \psi + \frac{1}{2} |\psi|^4 ] dx$, and the number of particles $N=\int_{-\infty}^{\infty} |\psi|^2 dx$. The two energy terms are the kinetic energy and the interaction energy, respectively.

The integrable GP equation has the following travelling dark soliton solution:
\begin{eqnarray}
\psi_d&=&\{\sqrt{\mu-v^2} \tanh[\sqrt{\mu-v^2}(x-v t)]+ i v\}  e^{-i \mu t}, \nonumber \\
\label{psid}
\end{eqnarray}
where $v$ is the dark soliton velocity, and $\mu$ is the chemical potential. It is immediately clear that the field energy $H$ is infinite, and we need to properly subtract the background energy to extract the energy of the spatially localized dark soliton.
The correct renormalized energy form of the dark soliton was suggested firstly by Y.S. Kivshar \textit{et al.} in 1994 \cite{Kivshar1} as:
\begin{eqnarray}
 E_{s} &=& \int_{-\infty}^{\infty} \left[ \frac{1}{2} |\partial_x\psi_d|^2+ \frac{1}{2} (|\psi_d|^2-\mu)^2 \right] dx, \\
 &=&\frac{4}{3} (\mu -v^2)^{3/2}.
 \label{dse}
\end{eqnarray}
This soliton energy form has been successfully applied to numerous studies of dark soliton dynamics in external potentials, yielding good agreement with experiments \cite{Kivshar2,Busch,Lewenstein,Konotop,Brazhnyi,Pelinovsky,Weller,darkex,Frantzeskakis}; see also the pertinent two-component generalization to the dark-bright soliton \cite{Busch2,Carr,Qu,Pitaevskii,zhaoliu,Ros}. The energy is also a key element for more exotic topics such as the negative mass of the dark soliton \cite{PNAS,Hurst,Serkin}. However, this crucial renormalized energy was suggested in a somewhat vague way, namely, it was stated that ``the soliton part of the total Hamiltonian (the system energy) may be defined as Eq.~(18)'' \cite{Kivshar1}.

Before introducing our method, it is instructive to examine a wrong but instructive derivation of the dark-soliton energy. At the same chemical potential $\mu$, the ground state is uniform $\psi_{g}=\sqrt{\mu}\exp(-i\mu t)$.
Considering simply the field energy difference, we find:
\begin{eqnarray}
E_{s}(\rm{wrong}) &=& H[\psi_d] - H[\psi_g], \\
&=& \int_{-\infty}^{\infty} \left[ \frac{1}{2} |\partial_x\psi_d|^2+ \frac{1}{2} |\psi_d|^4 \right] dx \nonumber \\
 &&- \int_{-\infty}^{\infty} \left[ \frac{1}{2} |\partial_x\psi_{g}|^2+ \frac{1}{2} |\psi_{g}|^4 \right] dx, \nonumber \\
 &=& -\frac{2}{3} (\mu +2v^2)\sqrt{\mu-v^2}.
\end{eqnarray}
The result is strikingly wrong, i.e., the excitation energy is negative, but the dark soliton is an excited state imprinted on the ground state. It is not hard to see that this negative energy stems from the ``lose of matter'' at the density dip. The gain in the kinetic energy is not sufficient to compensate the missing interaction energy. Our idea is exactly to correct this particle number difference as follows:
\begin{enumerate}
\item For a chosen interval around the dark soliton (Eq.~\ref{psid}) center $[-L, L]$, calculate the dark soliton field energy $H_L[\psi_d]$ and number of atoms $N_L[\psi_d]$. Note that the integral is restricted within the interval, and we assume that $L$ is much larger than the healing length $\xi$.
\item Calculate the ground state (under periodic boundary conditions) in the same interval such that it has the same number of atoms as the dark soliton, i.e., $N_L=N_L[\psi_d]=N_L[\psi_{gL}]$. The chemical potential and the field are denoted as $\mu_{g}(L)$ and $\psi_{gL} = \sqrt{\mu_{g}(L)} e^{-i \mu_{g}(L) t}$, respectively.
\item Evaluate the dark soliton energy, again for the finite interval, as $E_s(L)=H_L[\psi_d]-H_L[\psi_{gL}]$.
\item The dark soliton energy is finally extracted as $E_s = \lim\limits_{L\to\infty} E_s(L)$.
\end{enumerate}
In the next section, we present the detailed results. We will see that the final solution is remarkably identical to that of Eq.~\ref{dse}.

\section{Result}
\label{results}
%First, we examine again in more details of the wrong expression. As mentioned previously, the error is induced by the particle number difference between the dark soliton state and the ground state. The difference is $\Delta N= \int_{-\infty}^{+\infty} [\mu-|\psi_d|^2] dx= 2 \sqrt{\mu-v^2}$. The missing interaction energy is

First, we set the dark soliton center to $0$ for our integrals without loss of generality. The number of atoms for the dark soliton state is:
\begin{eqnarray}
N_L[\psi_d]&=& \int_{-L}^{L} |\psi_d|^2 dx, \nonumber \\
  &=& 2 \mu L - 2\sqrt{\mu-v^2} \tanh[\sqrt{\mu-v^2} L].
\end{eqnarray}
The particle number for the ground state is $N_{L}[\psi_{gL}]=2 L \mu_{g}(L) $. Requiring the ground state and the dark soliton state have the same number of atoms, the chemical potential of the ground state is calculated as:
\begin{eqnarray}
\mu_{g}(L) &=&\mu -\frac{\sqrt{\mu-v^2} \tanh[\sqrt{\mu-v^2} L]}{L}.
\end{eqnarray}
Note that $\lim\limits_{L\to\infty}\mu_g(L)=\mu$ as expected. But clearly we cannot simply set it to $\mu$ for the field directly, otherwise we would get the wrong expression again. Physically, the subtle difference means that the dark soliton excitation should lead to a small increase of the background density if the particle number is conserved. But the infinite size of the background hides this tiny variation.

Having the two fields in the finite interval in place, we are now ready to evaluate the tedious but straightforward dark soliton energy in the finite interval as:
\begin{eqnarray}
E_s(L) &=& H_L[\psi_d]-H_L[\psi_{gL}], \\
&=&\int_{-L}^{L} \left[ \frac{1}{2} |\partial_x \psi_d|^2+ \frac{1}{2} |\psi_d|^4 \right] dx\nonumber \\
 &&- \int_{-L}^{L} \left[ \frac{1}{2} |\partial_x\psi_{gL}|^2+ \frac{1}{2} |\psi_{gL}|^4 \right] dx, \nonumber \\
 &=& \frac{1}{6} (\mu-v^2)^{3/2} \sech^3[L'] (3 \sinh[L']+\sinh[3L']) \nonumber \\
 &&+ \frac{1} {3} (\mu-v^2)^{3/2} \sech^2[L'] \tanh[L'] (2+\cosh[2L']) \nonumber \\
 &&-\frac{1} {2L} (\mu-v^2)\sech^2[L'] \tanh[L'] \sinh[2L'],
\end{eqnarray}
where $L'=\sqrt{\mu-v^2}L$. Finally, taking the $L \rightarrow \infty$ limit,
\begin{eqnarray}
E_{s}&=&\lim\limits_{L\to\infty} E_{s} (L)=\frac{4}{3} (\mu -v^2)^{3/2}, \nonumber \\
&=&\int_{-\infty}^{\infty} \left[ \frac{1}{2} |\partial_x\psi_d|^2+ \frac{1}{2} (|\psi_d|^2-\mu)^2 \right] dx.
\end{eqnarray}
It is remarkable that these expressions are exactly what Y.S. Kivshar \textit{et al.} insightfully suggested. Here, we show that it is possible to derive these results analytically from the field energies by carefully keeping track of the density variation caused by the dark soliton excitation on top of the ground state.

Finally, we summarize here for completeness the ``derivation'' of the soliton energy in the grand canonical ensemble. In this setting, the dark soliton energy is defined from the difference of the ``grand canonical energy" or free energy $\Omega = H-\mu N$ \cite{Busch,Konotop}. Using the thermodynamic dark soliton state and the ground state of the same chemical potential $\mu$, the dark soliton energy can be rather straightforwardly calculated as:
\begin{eqnarray}
 E_{s}&=&\int_{-\infty}^{\infty} \left[ \frac{1}{2} |\partial_x\psi_d|^2+ \frac{1}{2} |\psi_d|^4-\mu |\psi_d|^2 \right] dx \nonumber \\
 &&- \int_{-\infty}^{\infty} \left[ \frac{1}{2} |\partial_x\psi_{g}|^2+ \frac{1}{2} |\psi_{g}|^4-\mu |\psi_{g}|^2 \right] dx, \\
 &=& \frac{4}{3} (\mu -v^2)^{3/2}.
\end{eqnarray}
While this approach is very mathematically efficient, the interpretation of this free energy difference as the dark soliton energy is a bit confusing. The system is actually not a grand canonical system, and the number of atoms is strictly conserved. This is also true experimentally, e.g., during the phase imprinting process or other excitation processes \cite{darkex1,darkex2}. The best way to understand this is perhaps to reasonably assume that these two ensembles here are equivalent. Our approach therefore in this sense is much more direct and intuitive. It is interesting, however, that all of these approaches yield the same dark soliton excitation energy $E_s=\frac{4}{3} (\mu -v^2)^{3/2}$.

\section{Conclusion}
\label{conclusion}
In this work, we revisited the fundamental problem of the dark soliton excitation energy and provided an alternative and intuitive approach to derive this energy. We use a limiting process to keep track of the tiny density variations of the background due to the dark soliton excitation, ensuring that the number of atoms is conserved. Our derivation uses only the elementary definition of the dark soliton excitation energy, field energy, and number of atoms, putting the well-known dark soliton excitation energy on a firm basis.

One possible future direction is to use this approach to investigate more exotic states such the vortex \cite{vortex1,vortex2,vortex3}. This is conceptually straightforward but appears to be analytically very challenging. But some approximations perhaps can be introduced. Research efforts along this line is currently in progress and will be presented in future publications.

\section*{Acknowledgements}
This work is supported by National Natural Science Foundation of China (Contact No. 11775176), Basic Research Program of Natural Science of Shaanxi Province (Grant No. 2018KJXX-094), The Key Innovative Research Team of Quantum Many-Body Theory and Quantum Control in Shaanxi Province (Grant No. 2017KCT-12), and the Major Basic Research Program of Natural Science of Shaanxi Province (Grant No. 2017ZDJC-32). W.W.~acknowledges support from the Fundamental Research Funds for the Central Universities, China.

\end{document}